\documentclass[12pt]{article}
\usepackage{sc3conf}
\usepackage{amsfonts}
\usepackage{epsfig}

\newcommand{\bee}{\begin{eqnarray}}
\newcommand{\eend}{\end{eqnarray}}
\newcommand{\rmd}{{\rm d}}
\newcommand{\rme}{{\rm e}}
\newcommand{\rmi}{{\rm i}}
\begin{document} \raggedbottom \hfill{Talk given at I.Ya.Pomeranchuk International Memorial
 Conference,\\
MIFI, Moscow, January 24-28, 2003 } \vspace{1.5cm}
\title{Superradiation in scattering of a Dirac particle off  a
point-like nucleus with $Z>$137 }

\authors{A.E.~Shabad}

\addresses{P.N.Lebedev Physics Institute, Russian Academy of Sciences,
Leninsky prospect 53, Moscow, Russia.\\
shabad@lpi.ru}

\maketitle

\begin{abstract}
The main concepts of the recently developed approach  to singular
problems of quantum mechanics are extended to the Dirac particle
in the Coulomb field of a point-like nucleus with its charge
$Z>$137. The reflection and transmission coefficients,  which
describe, respectively, the reflection of  electron by the
singularity and its falling onto it, are analytically calculated,
using the exact solutions of the Dirac equation. The
superradiation phenomenon is found. It suppresses the widely
discussed effect of spontaneous electron- positron pair creation.
\newpage
\end{abstract}
\section{Introduction}
In our previous publications \cite{shabad1, shabad2, shabad3} we
developed a new approach - called black-hole approach - to the
Schr$\ddot{\rm o}$dinger equation with attractive inverse-radius-
squared singularity in its potential, possessing the fall
-down
onto the center behavior. Within this approach the solutions,
oscillating near
 the point of singularity, are treated as free particles,
emitted or absorbed by the singularity exactly in the same sense,
as they are emitted or absorbed by the infinitely remote point -
we mean, income or outcome from/to infinity.

In the present work we apply this approach to the relativistic
problem of an electron placed in a Coulomb field of a point-like
nucleus with its charge $Z$ greater than 137. The corresponding
set of radial Dirac equations is known to possess in this case the
singularity of the same mathematical character as in the
Schr$\ddot{\rm o}$dinger problem above. Namely, the Coulomb
potential is squared in the Dirac equation and produces singular
attraction $(Z\alpha/r)^2$. Basing on the same concept and using
the known explicit solution of the
 Dirac equation, we are able to calculate the coefficients
 of reflection and transmission for electrons that income from infinity
and are then either reflected back or transmitted to the
singularity point.
 In the negative energy continuum $\varepsilon <-m$, where $m$ is electron
 mass, these coefficients are responsible for the spontaneous
electron-positron pair creation by the external field of the
nucleus. This effect has been studied in a more conventional
context (the basic references are \cite{zp}, \cite{gmr}), using an
introduction of the nucleus finite-size core, acting as a cutoff.
In the present
 approach the reflection and transmission coefficients
 are found analytically in a simple form, since they are not affected by
 a cutoff, whose presence is not meant at all. The result shows the
superradiation phenomenon, characteristic of the Kerr black holes,
when the reflected wave is more intense than the incident one. The
reflection coefficient is greater, than unity, the transmission
 coefficient being, correspondingly, negative. This fact forbids the
spontaneous pair creation owing to the Pauli principle applied within the 
Dirac sea picture.

The paper is organized as follows. In Section 2 we shortly outline
the concepts of the black-hole approach, addressing the Dirac equation
in the supercritical Coulomb field, with the reference to the original
work \cite{shabad3} for details. In Subsection 2.1 we formulate the
Kamke
generalized eigenvalue problem with respect to Z in what is called
sector III of confined states $-m<\varepsilon <m$,~$Z>137$; find the corresponding integration measure
in the orthogonality relations, that allows treating the solutions,
oscillating near the singularity, as free particles; perform the
 transformation (when $0<\varepsilon <m$) that reduces the Kamke
eigenvalue problem to the standard Liouville form and find the
corresponding effective potential, confining the particle near the
singularity. In Subsection 2.2 we analogously formulate the
effective barrier problem in what is called sector IV,
$\varepsilon^2>m^2$,~$Z>137$, where particles emitted by the
singularity or by the infinitely remote point are reflected back
or transmitted to, resp., the infinitely remote point or the
singularity. In Section 3 the reflection/transmission coefficients
are calculated. The recipe for this calculation is determined by
the concept, but is independent of any of the constructions in the
previous Section 2.
\section{Outline of the black-hole approach}
The radial Dirac equation in the Coulomb field is two-component
\cite{akhiezer}, \cite{blp}\bee\label{rad}\frac{\rmd G(r)}{\rmd r
}+\frac\kappa{r}G(r)-(\varepsilon
+m+\frac{Z\alpha}{r})F(r)=0,\nonumber\\\frac{\rmd F(r)}{\rmd r
}+(\varepsilon
-m+\frac{Z\alpha}{r})G(r)-\frac\kappa{r}F(r)=0.\eend Here the
spinor components of the radial wave function $G(r)$ and $F(r)$
correspond to $gr$ and $fr$ in notations of Ref.\cite{akhiezer},
resp., $\varepsilon$ and $m$ are the electron energy and mass,
$\alpha =1/137$ is the fine structure constant, $Z$ is the charge
of the nucleus, and $\kappa$ is the electron orbital
momentum\bee\label{kappa}
\kappa=-(l+1)\quad {\rm for}\quad j=l+\frac 1{2},\nonumber\\
\kappa=l\quad {\rm for}\quad j=l-\frac 1{2}.\eend In what follows
we confine ourselves to the lowest orbital state $\kappa=-1$.
Solutions to equation (\ref{rad}) are known in terms of confluent
hypergeometric functions. To illustrate the connection of the
problem (\ref{rad}) with the Schr$\ddot{\rm o}$dinger problem,
which possesses the fall-down onto the center property, let us for
example consider the combination \bee\label{comb} W(r)=
\sqrt{r}\left(\frac{G(r)}{\sqrt{1+\frac\varepsilon{m}}}-\frac{F(r)}
{\sqrt{1-\frac\varepsilon{m}}}\right).\eend  It satisfies the
second-order Schr$\ddot{\rm o}$dinger-like differential
equation\bee\label{sch} -\frac{\rmd^2W}{\rmd r
^2}-\left(\frac{\sqrt{m^2-\varepsilon^2}}{r} +\frac
1{4r^2}+\varepsilon^2-m^2\right)W=\left(\frac{Z^2\alpha^2-1}{r^2}+
\frac{2Z\alpha\varepsilon}{r}\right)W.\eend We collected all terms
containing the coupling $Z\alpha$ in the right-hand side. This
fixes the operator, whose eigenfunctions will be studied as
solutions to the generalized Kamke eigenvalue problem \cite{kamke}
with respect to $Z\alpha$. The term $Z\alpha/r$ in (\ref{sch})
corresponds to Coulomb attraction of an electron by a positively
charged nucleus, when $Z\alpha$ and $\varepsilon$ are
 positive, while the term $Z^2\alpha^2/r^2$ produces singularity, which
is attractive irrespective of the sign of $Z\alpha$. This property
of the relativistic problem is well-known. The fall-down onto the
center \cite{QM} occurs, provided that\bee\label{critical}
Z^2\alpha^2-1>0.\eend

Two fundamental solutions to equation (\ref{sch}) behave near the
origin $r\rightarrow0$ like\bee\label{like} r^{\pm\rmi\gamma+\frac
1{2}},\eend where $\gamma=\sqrt{Z^2\alpha^2-1}$.

Two fundamental solutions to equation (\ref{sch}) behave near
 the infinity $r\rightarrow\infty$ like\bee\label{asymp}
\rme^{\mp r\sqrt{m^2-\varepsilon^2}}~r^{\frac{\mp\rmi
Z\alpha\varepsilon} {\sqrt{\varepsilon^2-m^2}}\pm\frac 1{2}}.\eend

The usual Coulomb bound states of an electron lie within the strip
\bee\label{strip} -m<\varepsilon<m,\quad \alpha Z<1, \eend
corresponding to what was called sector I in \cite{shabad1,
shabad2,shabad3}. In this sector we choose the solution, which
decreases exponentially at infinity (the upper sign in
(\ref{asymp})). It is a linear combination of two solutions, out
of which one  decreases near $r=0$ ($\gamma $ is imaginary in
sector I),
 whereas the other increases in
accordance with (\ref{like}) and should be ruled out by the
 requirement of square-integrability. The nullification of the
 corresponding coefficient is possible for discrete values of
 $\varepsilon$, considered as functions of $\alpha Z$:
  $\varepsilon_{n_r}(\alpha Z),~n_r=0,1,2...$

On the contrary, in the strip\bee\label{stripIII} -m<\varepsilon
<m,\quad \alpha Z>1, \eend corresponding to what was called sector
III in \cite{shabad1, shabad2,shabad3}, the both behaviors
(\ref{like}) oscillate, and, in accord with the black-hole
approach of Refs. \cite{shabad1}, \cite{shabad2}, \cite{shabad3}
we treat them  as  particles, free near the origin. These
particles are emitted or absorbed by the singular center, but do
not escape to infinity due to (\ref{asymp}), with the upper sign
again. The corresponding states
 make a continuum, because the both oscillating solutions are
 acceptable. We call them confined states in order to distinguish
 from the bound states that also do not extend to infinity, but
 belong to discrete spectrum. The treatment of the solutions,
oscillating
near $r=0$, as responsible for free particles, is possible, since
the measure in the corresponding scalar product becomes infinite
in this point. Within the present context this will
 be seen later.

In sector II \bee\label{stripII} \varepsilon <-m\quad {\rm
or}\quad \varepsilon >m, \qquad \alpha Z<1\eend electron and
positron are elastically scattered from the Coulomb potential.

Finally, in sector IV \bee\label{stripIV} \varepsilon <-m\quad
{\rm or}\quad \varepsilon >m, \qquad \alpha Z>1\eend the wave
function oscillates both near the origin and at infinity, the
particles being free in the both regions. This corresponds to any
of the two inelastic scattering processes: An electron (or a
positron) incoming from infinity, is partially scattered back with
the probability, determined by the reflection
 coefficient, and partially penetrates to the origin, becoming a
free particle there. Symmetrically, the particle, emitted by the
 center, escapes to infinity with a certain probability, determined
 by the transmission coefficient, and partially is reflected back to
the center. We say that deconfinement processes go in sector IV.
\subsection{Sector III of confined states $-m<\varepsilon <m$,~
$Z\alpha >1$} Following \cite{shabad1, shabad2,shabad3}, equation
(\ref{rad}) or (\ref{sch}) should be in sector III  considered as
a generalized eigenvalue problem of Kamke \cite{kamke} ~ for the
parameter $Z$, while its solutions should be classified as
eigen-functions of the operator in the left-hand side of
(\ref{sch}). Following the standard procedure, we may multiply
eq.(\ref{sch}) by $W_1^*$, the solution to the equation, complex
conjugate to (\ref{sch}), written for a different eigen-value
$Z_1$. After subtracting the complex conjugate of the same product
with $Z_1$ and $Z$ interchanged, and integrating by parts, we
obtain \bee\label{orthog} -\left.\left(W_1^*\frac{\rmd W}{\rmd
r}-W\frac{\rmd W_1^*}{\rmd r}\right)\right|_{r_L}^\infty=\alpha
(Z-Z_1)\int_{r_L}^\infty W^*_1W\rmd\mu (r), \eend where
\bee\label{measure} \rmd \mu
(r)=\left(\frac{2\varepsilon}{r}+\frac{\alpha
(Z+Z_1)}{r^2}\right)\rmd r \eend is the integration measure with
which two eigenfunctions with different eigenvalues $Z_1\neq Z$
and the same $\varepsilon$ are orthogonal, provided that the
boundary term on the left-hand side in (\ref{orthog}) disappears.
This is guaranteed if the system is placed in a
box\bee\label{combox} r_L<r<\infty, \quad r_L\rightarrow 0, \eend
and the appropriate boundary condition \bee\label{zerro}
W(r_L)=0\eend is imposed at the wall of the box.

Let us make the transformation of the variable ( $r_0$ is a free
dimensional parameter)\bee\label{xi}\xi(r)=\int_{r_0}^r
\left(\frac 1{r^2}+\frac{2Z\alpha
\varepsilon}{\gamma^2r}\right)^\frac 1{2}\rmd r\eend and of the
wave function \bee\label{wave} \tilde{W}(\xi)= \Phi(r)W(r),\quad
{\rm where}\quad
\Phi(r)=\left(\frac{\gamma^2}{r^2}+\frac{2Z\alpha\varepsilon}{r}\right)
^\frac 1{4}.\eend This transformation is nonsingular when the
energy $\varepsilon > 0$ is positive, and reduces equation
(\ref{sch}) to the standard Liouville form \bee\label{standard1}
-\frac{\rmd^2\tilde{W}(\xi)}{\rmd
\xi^2}+U(\xi)\tilde{W}(\xi)=(\alpha^2 Z^2-1) \tilde{W}(\xi)\eend
with the effective potential\bee\label {potential1}U(\xi)=\frac{3+
\frac{2Z\alpha\varepsilon}{\gamma^2}r}{2(1+
\frac{2Z\alpha\varepsilon}
{\gamma^2}r)^2}-\frac{r\sqrt{m^2-\varepsilon^2}
+r^2(\varepsilon^2-m^2)+\frac 3{2}}{1+
\frac{2Z\alpha\varepsilon}{\gamma^2}r}.\eend It is understood that
$r$ is here the function of $\xi$, inverse to (\ref{xi}). The free
parameter $r_0$ enters the potential only through this function.
Unlike the nonrelativistic case of pure inverse square potential
($V=0 $ in Refs. \cite{shabad1, shabad2, shabad3}), where there is
no
 dimensional parameter originally present in the theory, so that it
is to be additionally introduced to define the problem, in the
present case under consideration we have at our disposal the
Compton length $m^{-1}$.

For negative $\varepsilon$ the potential (\ref{potential1})
becomes singular as a result of the singularity of the
transformation (\ref{xi}), (\ref{wave}) . 

 In the vicinity of the singular
point $r=0$ the transformation (\ref{xi}), (\ref{wave}) behaves as
\begin{equation}\label{21}
\xi=\ln\frac r{r_0},
\end{equation}
\begin{equation}\label{22}
\left(\frac{\gamma}{r}\right)^\frac 1{2}W(r)=\tilde{W}(\xi)
\end{equation}
The origin $r=0$ is mapped onto $\xi=-\infty$, and the infinitely
remote point $r=\infty$ to the point $\xi=\infty ~~(\xi\asymp
(2\sqrt{ 2Z\alpha\varepsilon}/\gamma)\sqrt{r})$. The asymptotic
behavior (\ref{like}) in the point $r=0$ becomes the oscillating
 asymptotic behavior at  $\xi\rightarrow -\infty$ (up to an
unessential constant factor)
\bee\label{1.3} \tilde{W}(\xi)\asymp\exp{(\pm\rmi
\xi\gamma)}
\end{eqnarray} with $\gamma=\sqrt{\alpha^2Z^2-1}$ playing the role of a
"momentum". The potential (\ref{potential1}) grows at
$\xi\rightarrow \infty $ as $((m^2-\varepsilon^2)\gamma^2/2\alpha
Z\varepsilon)\xi^2$ ( see Fig.1 ) and prevents the electron from
escaping to infinity.

\begin{figure}[htb]
  \begin{center}
   \includegraphics[bb = 0 0 405 210,
    scale=1]{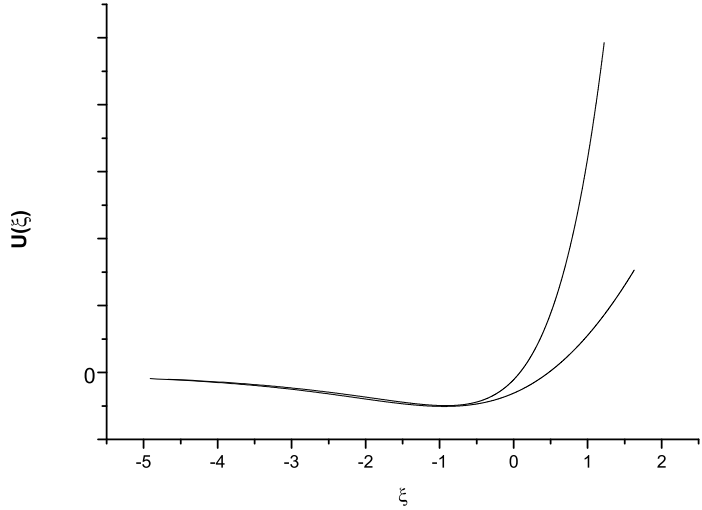}
    \caption{The confining potential (\ref{potential1}) in sector III
plotted for $Z\alpha=2$. The lower curve corresponds to the positive
energy value $\varepsilon=0.5 m$, and the upper one to
$\varepsilon =0.1 m$. The value $r_0=m^{-1}$ is taken for $r_0$} \label{Fig:1}
  \end{center}
\end{figure}

The box (\ref{combox}) is mapped to the box , restricted by the
limits \bee\label{box}  -\xi_L<\xi<\infty, \eend where
\bee\label{cutoff} r_L=r_0\exp{(-\xi_L) },\qquad
\xi_L\rightarrow\infty,~~~{\rm when}~~r_L\rightarrow 0.\eend The
boundary condition (\ref{zerro}) becomes\bee\label{zero} \quad
\tilde{W}(-\xi_L)=0 \eend in full analogy
 with the customary way of quantization of  particles, free at infinity. Eq.
(\ref{zero}) makes the Kamke problem self-adjoint and provides the
orthogonality and completeness of the wave functions (see
\cite{shabad3} for detail.)

When necessary,
the finite quantity $r_L$ may be considered as a cutoff, alternative
to the
nucleus core introduced in \cite{zp}.

As long as the box size $r_L$ is finite, the spectrum of
$Z\alpha$ is discrete in the domain $Z\alpha>1$ and is presented
by an equidistant series of vertical lines, numbered by
$n=0,1,2...$ from left to right in the plane
 with $y=\varepsilon $ for the ordinate, and
$x=Z\alpha >0$ for the abscissa, which fill the domain, restricted
by the borders of sector III, $i.e.$  $-m<y<m,~~1<x<\infty$. As
the spectral lines cannot intersect, these vertical lines must be
adjusted to the Coulomb bound states levels, which are described
in the same plane by a family of trajectories, numbered by the
radial quantum number $n_r\leq N$. $N$ is finite, provided that
$r_L$ is finite. These trajectories go in sector I, as $Z\alpha
$ grows, from the point $y=m,~~x=0$ to the
 border $x=1$ , separating
the two sectors. The lowest and leftmost trajectory $n_r=0$ joins
the leftmost vertical line $n=0$ near the point $y=0,~~x=1$ and is
continued by it down to $y=-m$. Other trajectories with
$n_r=1,2...N$ join with the vertical lines $n=1,2...N$ above that
point also to be continued down to $\varepsilon=-m$. The other
spectral lines with $n>N$ remain unaffected and go almost
vertically from $y=m$ to $y=-m$. When $r_L\rightarrow 0$, all the
spectral lines to the right of the border $x=1$ condense to make
the continuum of confined states. In this limit the bound state
trajectories approach that continuum at the border of sector III
at $x=Z\alpha =1$ and  at
discrete positions on the vertical axis,
corresponding to the discrete energy eigenvalues, numbered
 by $n_r$.

When $\xi_L\rightarrow\infty$, the norm of solutions of eq.
(\ref{standard1})\bee\label{norm} \int_{-\xi_L}
|\tilde{W}(\xi)|^2\rmd\xi =\gamma\int_{r_L} |W(r)|^2 \frac {\rmd
r}{r^2}  \eend diverges linearly with $\xi_L$.
 This makes the $\delta$-normalization of the wave functions possible.
Note the singularity of the volume element $\rmd r/r^2$ in the initial $
r$-space: the particle has enough volume near the origin to behave
itself freely near it.

 Such is the situation with sector III.

\subsection{Sector IV of inelastic scattering
$\alpha^2Z^2>1,~~\varepsilon^2>m^2$. Effective barrier}
 Excluding $F(r)$ from equation (\ref{rad}), one comes
\cite{zp} to
the equation 
for the upper component of the Dirac spinor $G(r)$\bee\label{G}
-\frac{\rmd^2G}{\rmd
r^2}-\frac{Z\alpha}{r(Z\alpha+r(m+\varepsilon))}\frac{\rmd G}{\rmd
r}-\left(\frac{2Z\alpha\varepsilon}{r}+\frac{-\kappa (
\varepsilon+m)}{r(Z\alpha+
r(m+\varepsilon))}\right)G=\nonumber\\
\left(\frac{Z^2\alpha^2-\kappa^2}{r^2}+\varepsilon^2-m^2
\right)G.\eend We placed the most singular  term in the r.-h.
side, as well the "kinetic energy" term $\varepsilon^2-m^2$.

The corresponding equation for the lower component $F(r)$ is
obtained from this by the substitution $G\rightarrow
F,~Z\rightarrow -Z,~\kappa\rightarrow
-\kappa,~\varepsilon\rightarrow -\varepsilon$.

Set $\kappa=-1$ and perform the transformation of the wave
function $G$, prescribed by the general theory of differential
equations \cite{kamke}, \bee\label{transformation2} \tilde{G} (\xi
)=\Phi (r)G(r),
\qquad\qquad\nonumber\\
\Phi (r)=\left(\frac{\alpha^2Z^2-1}{r^2}
+\varepsilon^2-m^2\right)^{\frac 1{4}}\left(\frac{r(\alpha
Z+r_0(m+\varepsilon ))}{r_0(\alpha Z+r(m+\varepsilon
))}\right)^{\frac 1{2}},\eend accompanied by the transformation of
the variable\bee\label{var}\xi (r)=(Z^2\alpha^2-1)^{-\frac
1{2}}\int_{r_0}^r\frac{\rmd
r}{r}\sqrt{\alpha^2Z^2-1+r^2(\varepsilon^2-m^2)}.\eend When
$r\rightarrow 0$ the new variable tends to $-\infty$ as $\ln r$.
When $r\rightarrow\infty$ the new variable tends to $\infty$ as
$\xi\rightarrow (\sqrt{\varepsilon^2-m^2}/\gamma)r$. After the
transformation (\ref{transformation2}), (\ref{var}) equation
(\ref{G}) acquires the standard Liouville
form\bee\label{standard2}-\frac{\rmd^2\tilde{G} (\xi )}{\rmd
\xi^2}+U(\xi)\tilde{G}(\xi )=(Z^2\alpha^2-1)\tilde{G}(\xi)\eend
with the effective potential\bee\label{potential2} U(\xi)=-\frac
1{(1+y^2)}\left(2\alpha Z\varepsilon r+\frac{(m+\varepsilon
)r}{\alpha Z+r(m+\varepsilon )}+\frac{\alpha Z}{4}\frac{\alpha
Z+4r(m+\varepsilon )}{(\alpha Z+r(m+\varepsilon ))
^2}\right.\nonumber\\\left.-\frac{6y^2+1}{4(1+y^2)^2}\right
),\quad y^2=r^2\frac{\varepsilon^2-m^2}{\alpha^2Z^2-1}.\eend If
the transformation of the coordinate is not used - only the wave
function is transformed - equation (\ref{standard2}),
(\ref{potential2}) is reduced to eqs. (2,11), (2,12), (2,13) of
Ref.\cite{zp}.

Equation (\ref{standard2}) should be supplemented by, $e.g.$,
periodic or antiperiodic boundary conditions in a  box $-\xi_L<\xi
<\xi_U$,~~$\xi_L\rightarrow \infty,~~\xi_U\rightarrow\infty$,
where $\xi_L$ is defined as (\ref{cutoff}) (\textit{cf.}
\cite{shabad3}). The nonrelativistic case, considered in
\cite{shabad3}, teaches us that, in sector IV, the complete and
orthogonal sets of solutions of the generalized Kamke eigenvalue
problem make families, for each of which a certain relation
between the two parameters $Z\alpha$ and $\varepsilon$ is fixed,
while the eigenfunctions within each family are labelled by the
remaining free parameter.
This relation for the present relativistic case remains unknown to
us.
\begin{figure}[htb]
  \begin{center}
   \includegraphics[bb = 0 0 405 210,
    scale=1]{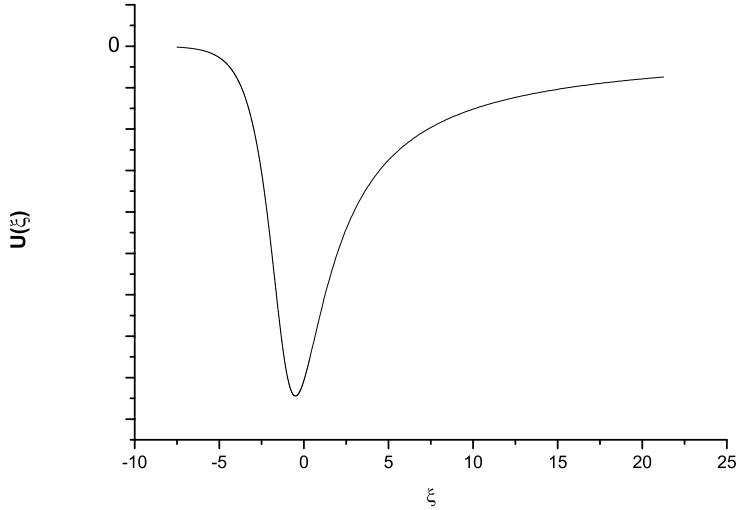}
    \caption{The barrier potential (\ref{potential2}) in the upper continuum.
 The curve is drawn for the positive energy $\varepsilon=2m$ and
the charge twice the critical value $\alpha Z=2$. The value
$r_0=m^{-1}$ is taken for $r_0$}.\label{Fig2}
  \end{center}
\end{figure}

In the upper continuum (a part of sector IV) $\varepsilon>m$ the
potential $U(\xi )$ (\ref{potential2}) is plotted against $\xi$ in
Fig.2. The most interesting situation appears in the lower
continuum $\varepsilon <-m$. The potential $U(\xi )$
(\ref{potential2}) is plotted for this case in Fig.3. It has the
barrier character and a singularity in the point $\xi_{\rm
sing}=\xi (r_{\rm sing})$, where \bee\label{sing}  r_{\rm
sing}=\frac{-\alpha Z}{m+\varepsilon }>0.\eend
\begin{figure}[htb]
  \begin{center}
   \includegraphics[bb = 0 0 405 210,
    scale=1]{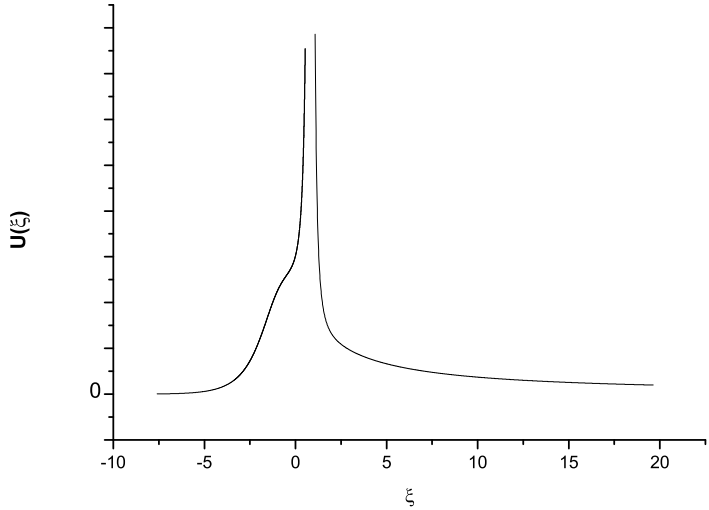}
    \caption{The barrier potential (\ref{potential2}) in the lower continuum.
 The curve is drawn for the negative energy $\varepsilon=-2m$ and
the charge twice the critical value $\alpha Z=2$. The value
$r_0=m^{-1}$ is taken for $r_0$}.\label{Fig3}
  \end{center}
\end{figure}
This singularity first appeared in equation (\ref{G}), resulted from
the exclusion of lower spinor component $F$ from equation (\ref{rad}),
 due to the zero in the coefficient in front of
$F$ in (\ref{rad}). Vice versa, one might exclude the upper spinor
 component $G$ from (\ref{rad}). Then one whould encounter the
 second-order differential equation like (\ref{G}) with the coefficients singular in
another point $r_{\rm sing}'=(Z\alpha/(m-\varepsilon )$, provided
that $\varepsilon <m$.

The singularity in the point (\ref{sing})  also  appears  in the part
of the potential $U_2$, "responsible for the  spin interaction",
eq.(2,13) of Ref.\cite{zp}, although is not discussed there. We
admit, that the superradiation phenomenon  discussed in the
next Section may be attributed to this singularity.
\section{Superradiation and the pair creation by the
 vacuum}
 The contents of this Subsection are  not directly dependent on the
  procedures developed above, since here we deal with
    known solutions, not with equations.

    In sector IV all solutions are meaningful, neither should be
ruled out. Let us consider the two-component solution to equations
(\ref{rad}), whose  behavior near the origin contains one
oscillating exponent\bee\label{GF} G(r)\asymp
(2pr)^{\rmi\sqrt{\alpha^2Z^2-1}}\left[\frac{\rmi\left(\frac{\alpha
Z\varepsilon}{p}+\sqrt{\alpha^2Z^2-1}\right)}{1-\frac{\rmi m
\alpha Z}{p}}
+1\right]\left(1+\frac\varepsilon{m}\right)^\frac 1{2},\nonumber\\
F(r)\asymp
(2pr)^{\rmi\sqrt{\alpha^2Z^2-1}}\left[-\frac{\left(\frac{\alpha
Z\varepsilon}{p}+\sqrt{\alpha^2Z^2-1}\right)}{1-\frac{\rmi m
\alpha Z}{p}} -\rmi\right]\left(\frac\varepsilon{m}-1\right)^\frac
1{2}, \eend where $p=\sqrt{\varepsilon^2-m^2}$. The same solution
behaves at  infinity $r\rightarrow\infty$ as\bee\label{FGinfty}
G(r)\asymp {\cal E}\rme^{-\rmi pr}(2pr)^{-\rmi\frac{\alpha
Z\varepsilon}{p}}\left(\frac\varepsilon{m}+1\right)^\frac
1{2}+{\cal G}\rme^{\rmi pr}(2pr)^{\rmi\frac{\alpha
Z\varepsilon}{p}}\left(\frac\varepsilon{m}+1\right)^\frac
1{2},\nonumber\\
 F(r)\asymp -\rmi{\cal E}\rme^{-\rmi pr}(2pr)^{-\rmi\frac{\alpha
Z\varepsilon}{p}}\left(\frac\varepsilon{m}-1\right)^\frac
1{2}+\rmi{\cal G}\rme^{\rmi pr}(2pr)^{\rmi\frac{\alpha
Z\varepsilon}{p}}\left(\frac\varepsilon{m}-1\right)^\frac
1{2}\eend The other  independent solution differs from this one by
the change of sign in the exponent in (\ref{GF}). (Correspondingly
the constant coefficients ${\cal E}$ and ${\cal G}$ in
(\ref{FGinfty}) become different.) The solution, described by the
asymptotic equations (\ref{GF}), (\ref{FGinfty}) is a wave,
falling from the infinity and reflected back to infinity, with a
transmitted wave, absorbed by the center. The coefficient $\cal G$
is responsible for transmission/absorbtion, and $\cal E$ for
reflection. As a whole, in sector IV, the inelastic  process is
described by two scattering phases and one mixing angle. These are
arranged in a unitary $2\times 2$ $S$-matrix, the reflection and
transmission coefficients being expressed in terms of these angles
\cite{shabad1, shabad2, shabad3}.

To find the absorbtion and reflection coefficients, note that the
conservation ${\rm div}{\bf j}=0$ of the current ${\bf
j}=\bar{\psi}{\bf\overrightarrow{\gamma}}\psi$ in the Dirac
equation implies that the flux\bee\label{flux} \rmi\int ({\bf
jn})r^2\rmd\Omega=FG^*-F^*G,\eend  where ${\bf n}={\bf r}/r$ and
$\Omega$ is the stereoscopic angle, be independent of $r$.
Eq.(\ref{flux}) is the Wronsky determinant for the set
(\ref{rad}). By equalizing the two expressions for the flux
(\ref{flux}) obtained near $r=0$ and $r=\infty$, we obtain the
current conservation property in the form of the
identity\bee\label{ident}1=\frac{|{\cal E}|^2}{|{\cal G}
|^2}+\frac{2\sqrt{ \alpha^2Z^2-1}}{|{\cal G} |^2\left(\frac{\alpha
Z\varepsilon}{\sqrt{\varepsilon^2-m^2}}-\sqrt{\alpha^2Z^2-1}\right)}.\eend
Here the second term in the r.-h. side is  the transmission
coefficient, and the first term the reflection coefficient. Using
the known explicit solution, whose asymptotic behavior is
(\ref{GF}), we find for the latter
coefficient:\bee\label{reflection} \mathbb{R}=\frac{|{\cal
E}|^2}{{|\cal G}|^2}=\rme^{-2\pi\gamma}\frac{\sinh
\pi(\zeta-\gamma)}{\sinh \pi(\zeta+\gamma)},\eend where
\bee\label{zeta} \zeta=\frac{\alpha
Z\varepsilon}{\sqrt{\varepsilon^2-m^2}},\quad
\gamma=\sqrt{\alpha^2Z^2-1}.\eend
\begin{figure}[htb]
  \begin{center}
   \includegraphics[bb = 0 0 405 210,
    scale=1]{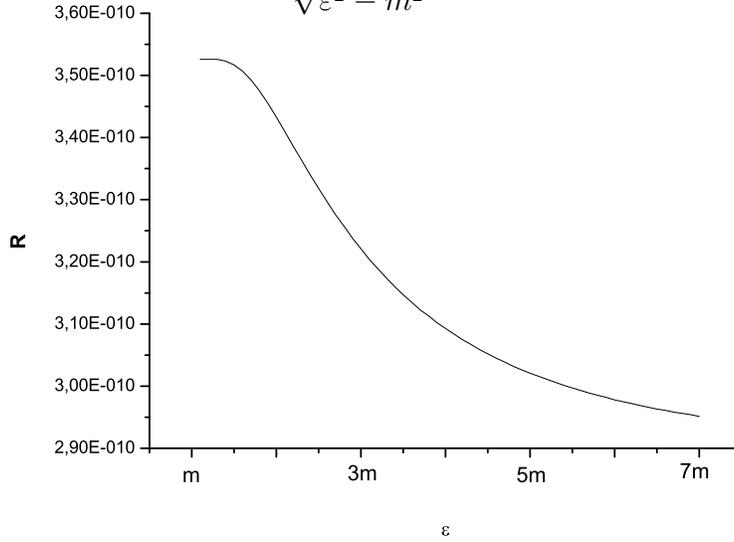}
    \caption{The reflection coefficient in the upper continuum
(for positive energies $\varepsilon > m$). The curve is drawn for
the charge twice the critical value $\alpha Z=2$.}\label{Fig4}
  \end{center}
\end{figure}

\begin{figure}[htb]
  \begin{center}
   \includegraphics[bb = 0 0 405 210,
    scale=1]{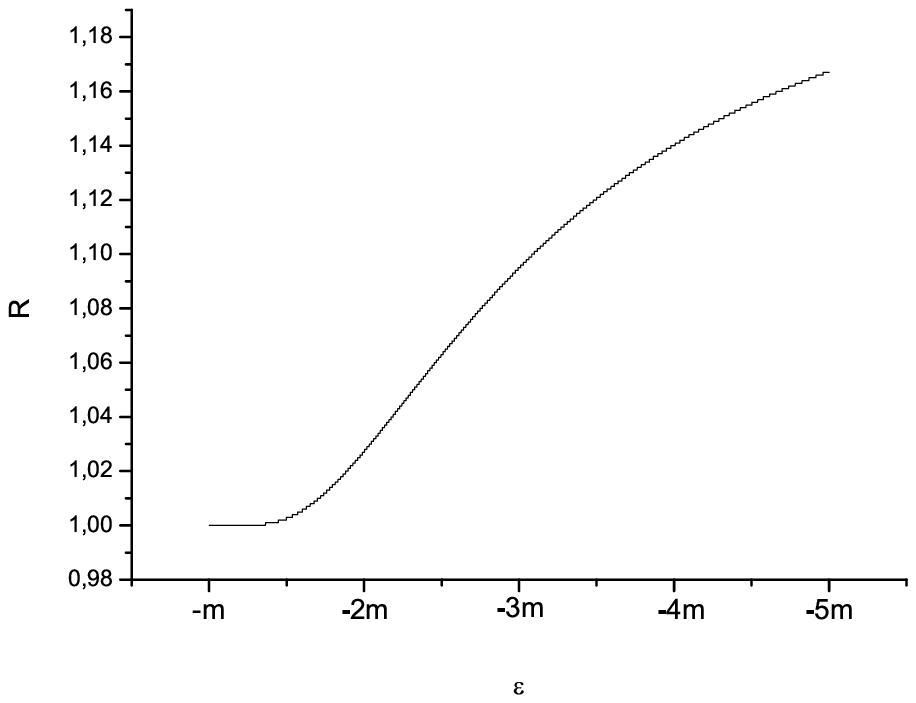}
    \caption{The reflection coefficient in the lower continuum
(for negative energies $\varepsilon < - m$). The curve is drawn
for the charge twice the critical value $\alpha
Z=2$.}\label{Fig.5}
  \end{center}
\end{figure}

The reflection coefficient (\ref{reflection}) is plotted against
the energy in Figs.4,5. In the positive kinetic energy domain
$\varepsilon >m$ the barrier, as shown in Fig.2, is negative
(attractive). Correspondingly, the reflection coefficient
(\ref{reflection}) is very small (see Fig.4). Its  maximum
achieved at the border $\varepsilon=m$ makes
 $\exp (-2\pi\gamma)$. The  Coulomb center acts as a strong absorber:
 it reflects next to nothing and absorbs almost everything.
  On the contrary, in the negative energy range
$\varepsilon <-m$ there is a  repulsing barrier singular in the
point (\ref{sing}) (\ref{potential2}), as shown in Fig.3.
Correspondingly, the reflection coefficient is  greater than
unity, as it follows from (\ref{reflection}) and is seen in Fig.5.
The phenomenon of $\mathbb{R}$ being greater than unity is called
superradiation. It is known for scattering off the Kerr (rotating)
black holes \cite{chandra} and is associated with the fact that
there appears singularity within the definition domain of the
differential equation. It is referred to as manifesting the
possibility to pump out energy from a black hole.

In our present context, the superradiation forbids the
electron-positron creation from the vacuum in the field of a
point-like nucleus with $Z\alpha >1$.

To explain this statement, let us first remind \cite{zp, gmr} that
this
 process may be traditionally understood within the second-nonquantized theory
as following. The nucleus is supplied with a core, which acts as a cutoff.
When, for sufficiently large charge $Z$, the electron level sinks into the
 negative continuum $\varepsilon <-m$, an electron taken from the filled
 Dirac sea can perform the under-barrier transition to become a
tightly bound state near the nucleus. At the same time, the hole
in the sea left by that electron, becomes a positron. In this way,
the pair, comprised of a free positron and a deeply bound electron
is produced from the vacuum by the strong cut-off
 Coulomb field. The probability of this process is determined
by the probability of the penetration through the barrier.

Our consideration of the nonregularized, point-like nucleus
modifies this idea in only one
 respect.
For us, the electron, after it passes through the barrier, becomes
a confined state, free near the singularity. The process is
described by the transition coefficient $\mathbb{T}=1-\mathbb{R}$.
Now, once the superradiation takes place, the number of electrons
reflected back into the Dirac sea, exceeds the number of those
taken from
 the sea. This is forbidden by the Pauli principle, since all vacancies in the sea are occupied.
In this way the superradiation supports the vacuum stability.
\section{Conclusion}
The mathematical basis for application of the present black-hole
approach to the relativistic problem of an electron in the
overcritical Coulomb field is not as firmly established as it was
in the case of the Schr$\ddot {\rm o}$dinger equation
\cite{shabad3}. Lacking  is a discription of families of sets of
complete orthogonal solutions to the Kamke eigenvalue problem in
sector IV, where inelastic processes occur. Nonetheless, the
adjustment of the black-hole approach into relativistic context
undertaken in Section 2 demonstrates its conceptual relevance,
sufficient for calculating, in Section 3, the reflection and
transmission coefficients, responsible for spontaneous pair
creation. The establishing of superradiation in electron
scattering off the Coulomb center with $Z>137$, which suppresses
the pair creation is independent of the way the Kamke problem is
posed and solved.
\section*{Acknowledgements}
The author acknowledges the financial support  of Russian Foundation for
Fundamental Research ( RFFI 02-02-16944 ) and the President of
Russia Program for Support of Leading Scientific Schools
(LSS-1578.2003.2).

\end{document}